\documentclass[a4paper, 
12pt, 
]{article} 

\usepackage[pdfstartview=FitB]{hyperref}
\usepackage{natbib}
\usepackage{geometry}
\usepackage{subcaption} 
\usepackage{amsmath,amssymb,amsthm}
\usepackage{graphicx}
\usepackage{todonotes}
\usepackage[doublespacing]{setspace}
\usepackage{float}
\usepackage{algorithm} 
\usepackage{algpseudocode} 

\usepackage{algorithm} 

\usepackage{algpseudocode} 

\makeatletter

\newcommand{\algmargin}{\the\ALG@thistlm}

\makeatother

\newlength{\whilewidth}

\algnewcommand{\parState}[1]{\State%
\parbox[t]{\dimexpr\linewidth-\algmargin}{\strut #1\strut}}

\theoremstyle{plain}

\theoremstyle{definition}

\title{Big Data meets Causal Survey Research: Understanding Nonresponse in the Recruitment of a Mixed-mode Online Panel}
\author{Barbara Felderer \thanks{GESIS Leibniz Institute for the Social Sciences \newline Correspondence: Barbara Felderer, GESIS Leibniz Institute for the Social Sciences. B2,1 68159 Mannheim. Email: barbara.felderer@gesis.org } \and Jannis Kueck \thanks{University of Hamburg} \and Martin Spindler \footnotemark[2]}
\begin{document}

\date{}
\maketitle

\begin{abstract}
Survey scientists increasingly face the problem of high-dimensionality in their research as digitization makes it much easier to construct high-dimensional (or “big”) data sets through tools such as online surveys and mobile applications. Machine learning methods are able to handle such data, and they have been successfully applied to solve \emph{predictive} problems. However, in many situations, survey statisticians want to learn about \emph{causal} relationships to draw conclusions and be able to transfer the findings of one survey to another. Standard machine learning methods provide biased estimates of such relationships. We introduce into survey statistics the double machine learning approach, which gives approximately unbiased estimators of causal parameters, and show how it can be used to analyze survey nonresponse in a high-dimensional panel setting.
\end{abstract}
\section{Introduction}\label{intro}

A key attribute of “big data” is  the large volume of data that is collected or generated, often for the purpose of statistical analysis (for further attributes see, for example, \citet{10.1093/poq/nfv039}). When a large number of observed characteristics are available for only a limited number of observations, however, the high-dimensionality of the data sets poses challenges. Moreover, big data comes in a variety of forms, including many sorts of paradata \citep{kreuter2013improving} such as call records, time stamps or device-type and questionnaire-navigation data from online surveys \citep{callegaro2013paradata}, as well as sensor data from mobile surveys \citep{struminskaya2020augmenting} and data from outside sources that can augment survey data and be linked to persons or population groups by unique personal or group identifiers. These outside data contain, for example, administrative records (cf. \citet{durrant2009multilevel} for nonresponse analysis), data from social media (an extensive discussion on the role of social media in public opinion research can be found in \citet{murphy2014social}) or regional information (e.g., \citet{feddersen2016subjective} study the impact of weather and climate on self-reported life satisfaction). Increasingly, the field of survey analysis is facing the challenges posed by high-dimensional data sets. Long-lasting panel surveys produce big data, for example, by collecting large numbers of variables over many panel waves. Some frequently used methods cannot be employed with big data sets that have comparatively few observations and numerous variables. To deal with problems of high dimensionality, machine learning methods have found their way in survey research modeling (see, for example, \citet{Buskirk2018_intro}, \citet{Buskirk2018}, \citet{Kirchner2018}, \citet{Adam2018} and \citet{kern2019tree} for introductions of the use of machine learning techniques with survey methodological questions).

Generally speaking, there are two main kinds of statistical modeling: causal inference (also known as explanatory analysis) and predictive modeling. Both have their own model-building logic and evaluation tools \citep{breiman2001statistical}. As \citet{shmueli2010explain} states, high predictive power does not necessarily imply high explanatory power, so different tools should be used to explain and to predict. The aim of prediction models is to predict the dependent variable $y$ for individuals who were not among those used to build the model. The best model is found, for example, by minimizing the out-of-sample mean squared error (MSE). Modern machine learning methods have been highly successful at building predictive models. In contrast to predictive modeling, causal inference entails learning the effect of a particular variable on the dependent variable $y$ while holding all other variables constant. Being able to draw ceteris paribus conclusions in this manner, researchers can think about interventions (i.e., changing $x$ will affect $y$ in a known way) and use this to design future studies. Applying modern machine learning methods to gain explanatory insights, however, is more challenging than building predictive models because machine learning methods inevitably introduce some bias in the estimation \citep{BCH2014}. In the recent years, progress has been made in applying machine learning to causal inference, and tools for doing so, such as the double machine learning framework, have been developed. In this paper, we demonstrate how survey statistics can benefit from these methods, obtaining insights into dealing with high-dimensional survey data sets by applying the double machine learning method to learn about nonresponse in the recruitment of the GESIS panel.

Survey nonresponse is arguably one of the chief problems in survey research \citep{kreuter2013facing} and many decades of study have been invested in developing methods to explain and thereby prevent or adjust for it (for recent examples see \citet{durrant2009multilevel, rossmann2016using}). With the rise of big data and the increasing number of variables being considered, one of the more recent methods is machine learning. Multiple studies have demonstrated its usefulness in this context: For example, \citet{kern2019tree} show that regression trees can effectively be used to predict nonresponse in the German Socio-Economic Panel; \citet{phipps2012analyzing} use trees to analyze nonresponse in an establishment panel and \citet{buskirk2015finding} use random forest classification models and random forest relative class frequency models to predict response propensities in a simulation study. Other examples are
\citet{Signorino2018}, who employ adaptive lasso to predict nonresponse in the National Health Interview Survey; \citet{earp2014modeling}, who use an ensemble of classification trees to predict nonresponse in an establishment survey's subsequent wave; \citet{kern2019longitudinal}, who apply different machine learning methods to predict nonresponse using information from multiple waves of the GESIS panel; and \citet{Zinn2020}, who use Bayesian additive regression trees to predict temporary and permanent dropout in an event history analysis in the German National Educational Panel Study. Finally, \citet{liu2020using} compare the use of random forests, support vector machines and lasso regression to predict response in the second interview of the Surveys of Consumers national telephone survey. 

As mentioned above, one must be careful when the results produced by machine learning algorithms are interpreted beyond predictions. While nonresponse prediction can be seen as a goal in its own right, one must be clear about its limitations: the effects of the control variables cannot be interpreted because machine learning algorithms -- when applied directly -- inevitably introduce bias, and thus no understanding of any causal effects of explanatory variables on the dependent variable of interest can be gained. Nonresponse prediction models help to identify individuals who are most likely to drop out but do not allow us to understand the driving factors, which are, however,  key to identifying and developing prevention strategies \citep{lynn2017standardised}. 

In this paper, we use machine learning methods not only to predict nonresponse, but to analyze explanatory factors in a high-dimensional setting for survey statistics. Recently, double machine learning techniques to deal with high dimensions and to deliver unbiased estimates have been developed (cf. \citet{CHS:2016, belloni2017program, chernozhukov2018double}). We give an introduction to the double machine learning approach and show how double lasso can be applied to explain nonresponse in the welcome survey of the GESIS panel. Using our causal machine learning approach, we find that nonresponse is affected by respondents' socio-demographic characteristics, and interviewers' ratings of both the respondents' cooperativeness during the interview and the respondents' likelihood to participate in the welcome survey. Socio-demographics are additionally found to interact with the chosen mode of participation. Our findings can help survey researchers who design and implement panel surveys to develop targeted strategies to prevent nonresponse.

The rest of the paper is structured as follows: In Section  \ref{secb} we introduce the basic principles of double machine learning, focusing on double selection for logistic regression models. In Section  \ref{application}, we describe an application for nonresponse modeling in the GESIS panel. We conclude with a discussion in Section \ref{sec4}.

\section{Double Machine Learning}\label{secb}

Machine learning methods have been developed mostly for prediction problems, which are based on finding correlations among variables. Often the machine learning algorithm is considered to be a black box that delivers acceptable forecast accuracy but in which the interaction of the variables is not understood. In many situations, however, scientists and practitioners are interested in learning the effect of certain variables, often called treatment variables, on one or more dependent variables, holding all other factors constant. This is more challenging than building a predictive model because here the black box must be opened and the inner mechanism learned.

Almost all machine learning methods, like lasso, lead to biased estimates of causal relationships and hence invalid inference results, despite their predictive power \citep{BCH2014}. In recent years, frameworks for valid post-selection inference have been developed. The double machine learning framework we present in the following section allows for such valid inference and hence learning about parameters and explanatory variables in a high-dimensional setting.

\subsection{Basic Setting and Idea behind Double Machine Learning}

In this section, we would like to introduce the basic ideas behind double machine learning. 
The goal is to estimate the treatment effect $\alpha_0$ of a treatment variable $D$ on the dependent variable $Y$ in a high-dimensional setting, namely

\[
Y = \gamma + \alpha_0 D + g(X) + \varepsilon,\quad\mathbb{E}(\varepsilon|D,X)=0,
\]
where $\gamma$ is the intercept and $g(\cdot)$ a function of the control variables. The set of control variables $X=(X_1,\ldots,X_p)$ might be high-dimensional. The most common case, which we will focus on here, is a linear approximation $g(X)= \beta_1 X_1 + \ldots \beta_p X_p$, with $\beta=(\beta_1,\ldots, \beta_p)$ as nuisance parameters.

Our goal is to perform valid inference on the treatment parameter $\alpha$ in a high-dimensional setting, i.e. the number of variables $p$ might be larger than the number of observations $n$. The function $g$, or in the linear case the parameter vector $\beta$, are considered nuisance parameters and are not part of the model interpretation.\\
For ease of exposition, we consider the case of one treatment variable here, but several treatment variables can just as easily be considered and the effects estimated at the same time. If the number of variables or hypotheses to test becomes large, methods from simultaneous inference may be applied (for a survey on recent developments, we refer to \citet{Bach2018}).

In a high-dimensional setting, standard ordinary least squares (OLS) estimation is not appropriate because of overfitting, which leads to poor estimates and forecasts. A naive approach often employed by empirical researchers is to use lasso to select the relevant regressors first and then to conduct an OLS regression of the dependent variable $Y$ on the treatment variable $D$ and the selected regressors from the lasso regression. This procedure, however, leads to biased results because lasso can fail to select variables that are strongly correlated with the treatment variable but only weakly correlated with the dependent variable. While this does not harm the predictive performance of the lasso, it leads to omitted variable bias \citep{BelloniChernozhukovHansen2011}, which biases the inference results. To correct for this problem, a de-biased lasso/double machine learning approach was introduced by \citet{chernozhukov2018double}. To understand this approach, we introduce an auxiliary equation for the treatment variable, as follows:
$$ D = \gamma_1 X_1 + \ldots + \gamma_p X_p + \nu.$$

The idea of double machine learning is to run a lasso regression of the auxiliary model to identify which variables create the omitted variable bias in the first step and subsequently include them in the final regression step. It can be shown that this approach leads to estimates of the target parameter that are asymptotically normally distributed (allowing valid post-selection inference). Introducing this auxiliary regression step and including omitted variables in the final regression implicitly creates a moment condition for the target parameter that fulfills the so-called Neyman orthogonality property. This means that the derivative with regard to the nuisance parameter of the corresponding score function is equal to zero at the true parameter values. Intuitively, we can see that small errors in the estimation of the nuisance parameter, as they occur under lasso, do not have a first-order effect on the treatment parameter. Despite selection errors in the confounders, valid results are achieved.

\subsection{Double Selection for Logit Models}

In many survey applications, the dependent outcome variable is binary, and for binary outcome variables, logistic regression is often the approach of choice. For logistic regression, the same arguments as outlined above apply when modern machine learning methods such as lasso are used to select variables and estimate the coefficients. To enable valid post-selection inference for the logistic regression, the double machine learning approach has to be modified appropriately (cf. \citet{belloni2013honest}).


\subsubsection{Logistic Regression}

In the logistic regression, a binary dependent variable $Y$ relates to a scalar treatment $D$ of interest and a $p$-dimensional control $X$ through a link function $G$

\[
E[Y|X,D] =G(D \alpha_0 +X' \beta_0).
\]

For logistic regression, the link function is given by $G(t) = \exp(t)/\{1+\exp(t)\}$. We aim to perform statistical inference on the coefficient $\alpha_0$, which represents the impact of the treatment on the dependent variable through the link function. Estimation is usually based on the (negative) log-likelihood function associated with the logistic link function, as follows:
\begin{align*}
\Lambda_i(\alpha,\beta) = \log \{1 + \exp(D_i \alpha + X'_i \beta)\} Y_i (D_i \alpha + X'_i \beta).
\end{align*}

For estimation in a high-dimensional setting, an $\ell_1$-penalty term, $||(\alpha,\beta)||_1= |\alpha|+\sum_{j=1}^{p}|\beta_j|$, is added to the minimization problem. The lasso logistic regression estimator is given by:
\begin{align*}
\left(\hat{\alpha}, \hat{\beta}\right) &\in \arg \min_{\alpha, \beta} \mathbb{E}_n[\Lambda_i(\alpha,\beta)] +\lambda/n ||(\alpha, \beta)||_1,
\end{align*}
where $\lambda$ is the penalty level and $\mathbb{E}_n$ denotes the empirical mean. As discussed in the section above, inference on the treatment parameter $\alpha_0$ is challenging and requires a modified estimation method, e.g., the de-biasing lasso estimator, based on a modified moment condition. The algorithm for the de-biased estimation of the treatment parameter $\alpha_0$ is presented in Algorithm \ref{dmlalg} in Appendix \ref{sec_dml}.

\section{Application: Nonresponse modeling for the GESIS panel}\label{application}
To illustrate the double machine learning lasso, we apply the technique to model nonresponse in the 2013 recruitment to the GESIS panel.

\subsection{Nonresponse in Panel Recruitment}
Recruitment to a probability-based panel is arguably the most important and most expensive part of the panel life-cycle. The recruited sample needs to represent the target population in order for valid inferences to be drawn for that population, and the sample size needs to be large enough to obtain precise estimates. The recruitment process usually includes several steps: contacting sampled cases and inviting them to a first recruitment survey, conducting this recruitment interview and, often during it, obtaining consent to proceed in the panel.  Consenting respondents are then invited to a welcome survey (or profile survey), and those who complete it are considered to be panel members. The panel members are then surveyed on a regular basis.

Even if the regular panel waves are conducted in a self-administered mode (e.g., by mail questionnaire and/or online), it is common to approach sampled persons and conduct the recruitment interview in an interviewer-administered (face-to-face or telephone) mode \citep{blom2016comparison}. Respondents to the recruitment survey are then asked to proceed with the subsequent survey using cost-saving self-administered modes. This, however, includes a switch in response mode that may be subject to systematic nonresponse. 

For our application, we choose nonresponse in the first interview after this switch of modes. We consider this stage to be very important for several reasons: First, this is when a large number of respondents to the recruitment survey are usually lost (for nonresponse rates in four large-scale scientific (mixed-mode) online surveys, see \citet{blom2016comparison}), and there is need to understand nonresponse in order to prevent it, i.e., by tackling likely nonresponse through targeted invitations \citep{lynn2020}. Second, nonresponse among respondents to the face-to-face interview is costly if we consider that they have completed the cost- and labour-intensive personal interview and are no longer available to take part in the less expensive self-administered part of the panel. In addition, refreshment samples are usually planned for panels once the number of respondents has fallen below a certain minimum number. Starting with a smaller sample means that costly new recruitment is needed sooner. Third, nonresponse can introduce bias to the panel. If the respondents are not lost at random, analyses of panel data can be severely biased. 
 
While a number of studies have been published on panel attrition, e.g., nonresponse to individual panel waves or dropout from the panel, the literature about nonresponse at the recruitment stages is surprisingly scarce. \citet{Sakshaug2020} analyze total recruitment error, which they define as error from initial nonresponse plus error from non-consent to be contacted again. In their comparison of a self-administered (mail/web) and CAPI recruitment, they find, for both modes, nonresponse bias to be larger than non-consent bias and total recruitment bias to be similar in both groups:  both recruited samples overrepresent older and more educated population groups, currently employed persons and higher-wage groups and underrepresent foreign-born persons. For the GEISIS recruitment panel, \citet{bosnjak2018establishing} find age, citizenship, marital status, household size, place of birth, education and household income to be distributed differently among the sample of respondents compared to the general population, with the differences tending to be larger for the welcome survey. 

In contrast to the initial recruitment survey, in which usually only a few variables from the sample frame are available, the recruitment interview usually generates a lot of information on the respondent, facilitating the study of nonresponse in the welcome survey. In addition to basic-sociodemographic information, the recruitment survey often includes information on attitudes and survey experience. In interviewer-administered surveys, the interviewers often provide information about the interview situation and their expectations of the respondents’ future participation in the panel. In particular, interviewers’ ratings of a respondent’s propensity to participate in a future survey, as well as ratings of cooperativeness and enjoyment, have been found to improve nonresponse models (see for example \citet{sinibaldi2015using, Plewis2017can}). Understanding the nonresponse process better can help to identify measures to address the problem, for example through targeted invitations \citep{lynn2020}.

While having a rich set of factors that potentially influence nonresponse is very helpful to understanding the nonresponse decision, it poses a challenge to nonresponse modeling. Indeed, including a large number of variables, possibly split into multiple dummy variables, and interactions requires big data solutions.

\subsection{The GESIS panel data}
The GESIS panel \citep{bosnjak2018establishing} is a probability-based, mixed-mode online and postal mail panel conducted bimonthly by GESIS -- Leibniz Institute for the Social Sciences in Mannheim, Germany. The first cohort of the GESIS panel was recruited in 2013 and refreshment samples were recruited in 2016 and 2018.
Recruitment to the GESIS panel in 2013 was based on a random sample of $21,870$ German-speaking residents of Germany aged $18$ to $70$ during the year of recruitment. In the first step, all sampled cases were invited to participate in a face-to-face recruitment survey. During this survey, respondents were asked for their consent to be invited to the GESIS panel by means of the self-administered online mode or the paper and pencil mode. Consenting respondents were then invited to participate in the welcome survey in the mode of their choice. Only after completing the welcome survey were respondents considered to be GESIS panel members.

In our study, we analyze nonresponse in the 2013 welcome survey among consenting respondents. We use data from the GESIS panel registration survey in 2013 \citep{ZA5665} to model nonresponse (or drop-out) (yes/no) in the subsequent welcome survey. 
In total, $7,599$ persons participated in the face-to-face registration survey, of whom $6,210$ agreed to being invited to the welcome survey and participating in the GESIS panel. Of these individuals, $4,938$ responded to the welcome survey and thus became regular panel members (dropout rate: $20.5\%$).

For our final sample, we drop $302$ observations with missing information, leaving us with $5,908$ respondents from the recruitment survey, of whom $4,720$ completed the welcome survey (dropout rate: $20.1\%$). In our analysis, we use 63 initial regressors representing information collected in the recruitment interview. This includes socio-demographic characteristics of the individuals and their cooperativeness throughout the interview. The variables we include in the analysis are listed in Table \ref{listofvars}. We transform categorical variables into level-wise dummies and add interaction terms of the regressors. This ultimately leads to a high-dimensional logit model with a total of $329$ regressors:

\begin{align*}
E[Y|X,D] = \frac{\exp(D \alpha_0 +X' \beta_0)}{1+\exp(D \alpha_0 +X' \beta_0)}.
\end{align*}

The binary dependent variable $Y$ indicates nonresponse to the welcome survey. The regressors split up into $303$ control variables $X$ and $26$ treatment variables $D$.
For the treatment variables, we choose key socio-demographics, the mode the respondents chose for the welcome interview (paper-and-pencil or online questionnaire) and interviewer ratings collected in the recruitment survey. The interviewer ratings include three cooperativeness ratings and one rating of individuals' willingness to participate in the welcome interview. The questions are:

\begin{itemize}
\item How would you rate the respondent’s willingness to answer the questions? (answer categories: good, moderate, low, good in the beginning but got worse, low in the beginning but got better)
\item How difficult or easy was it to persuade the respondent to take part in the interview? (answer categories: very difficult, rather difficult, rather easy, very easy)
\item How difficult or easy was it to persuade the respondent to take part in the follow-up interview? (answer categories: very difficult, rather difficult, rather easy, very easy)
\item How likely is it that the respondent will take part in the first online- or paper questionnaire? (answer categories: very likely, rather likely, rather unlikely, very unlikely)
\end{itemize}

We combine sparse categories with other categories for our analysis. We recode the answer categories into good vs. bad/all other categories for  ``willingness to answer the questions'' and combine very difficult and difficult for the two questions on the difficulty of persuading respondents to take part in the interview and follow-up interview. For the rating of the likelihood of response to the first online or paper questionnaire, we combine rather unlikely and very unlikely. With regard to sociodemographics, we include age, gender, highest educational degree, country of birth and living situation. We generate the living situation variable from information on marital status, partnership and living in a shared household leading to the five categories: \emph{no partner}; \emph{partner, not in household}; \emph{partner, in household}; \emph{married, living together}; \emph{married, living apart}. An overview of the coding for all treatment variables can be found in Table \ref{treatment_coding} in the Appendix. We include interactions of the choice of mode for the welcome survey with age, education and living situation to account for differential effects of the choice of mode on nonresponse.

\begin{table}[h]
\centering
\begin{tabular}{lrrrr}
\hline
   Treatment variables \\
\hline
gender, age, nationality, education, living situation, invitation mode,\\
willingness to answer the questions, willingness to participate in the interview, \\
willingness to participate in the panel, probability of participating in the survey\\
\hline 
   Control variables \\
\hline 
migration, employment status, occupational group, life satisfaction, leisure time, \\
country of birth, internet use, technical affinity, survey experience, \\
household size, number of children, income, incentive point, \\
invitation hesitance, interview intervention\\ \\
\hline 
\end{tabular} 
\caption{Extract of Regressors} 
\label{listofvars}
\end{table}

\subsection{Results}\label{results}
\begin{figure}[h]
\centering
\includegraphics[scale=1]{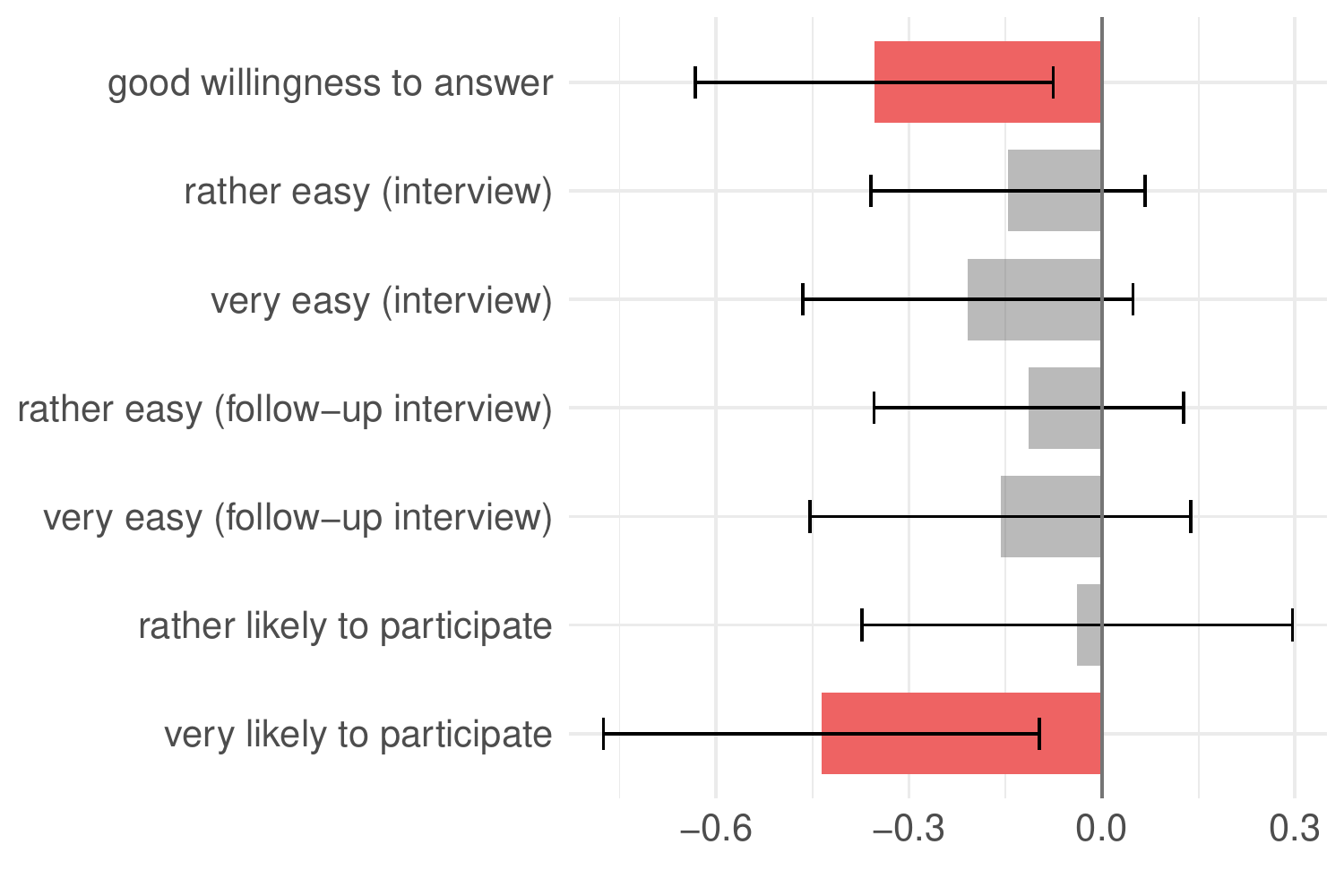}
\caption{Regression coefficients of the interviewer ratings in the logistic regression model.}
\label{fig_interview}
\end{figure}

In this section, we present the results of our double machine learning approach to the inferential analysis of nonresponse in the GESIS panel. The results of the double lasso for logistic regression are visualized in Figures \ref{fig_interview} to \ref{fig_online}, and a regression table can be found in Table \ref{treatment_effects} in the Appendix. 
We start with the interpretation of the interviewer ratings. The estimated coefficients of the interviewer ratings from the logistic regression together with the corresponding confidence intervals are displayed in Figure \ref{fig_interview}.

\paragraph{Cooperativeness} We find that the interviewer observation of respondents’ willingness to answer the survey questions in the recruitment survey had a highly significant negative effect on survey nonresponse. Respondents who were rated as having good willingness to respond to the recruitment survey dropped out of the survey after the recruitment stage to a lesser extent than respondents who were rated as having low willingness. We do not find significant effects for the ease of persuading respondents to participate in the interview nor for the ease of persuading respondents to consent to be contacted again for the follow-up interview. The effects however tend in the same direction as the observed willingness to answer the questions: respondents who were rated as being rather easy or very easy to persuade were less likely to drop-out.  

\paragraph{Rated likelihood of participation} Respondents who were rated as being rather or very likely to participate in the welcome survey dropped out after the recruitment survey to a lesser extent than did those who were rated as being rather unlikely or very unlikely to participate. We, however, find that the only significant effect in this regard is for “very likely” category.

\paragraph{Socio-demographics and chosen survey mode} Next, we discuss the effects of socio-demographics and chosen survey mode for the welcome survey. 
The results are found in Figures \ref{fig_socio} and \ref{fig_online} .

\begin{figure}[h]
\centering
\includegraphics[scale=1]{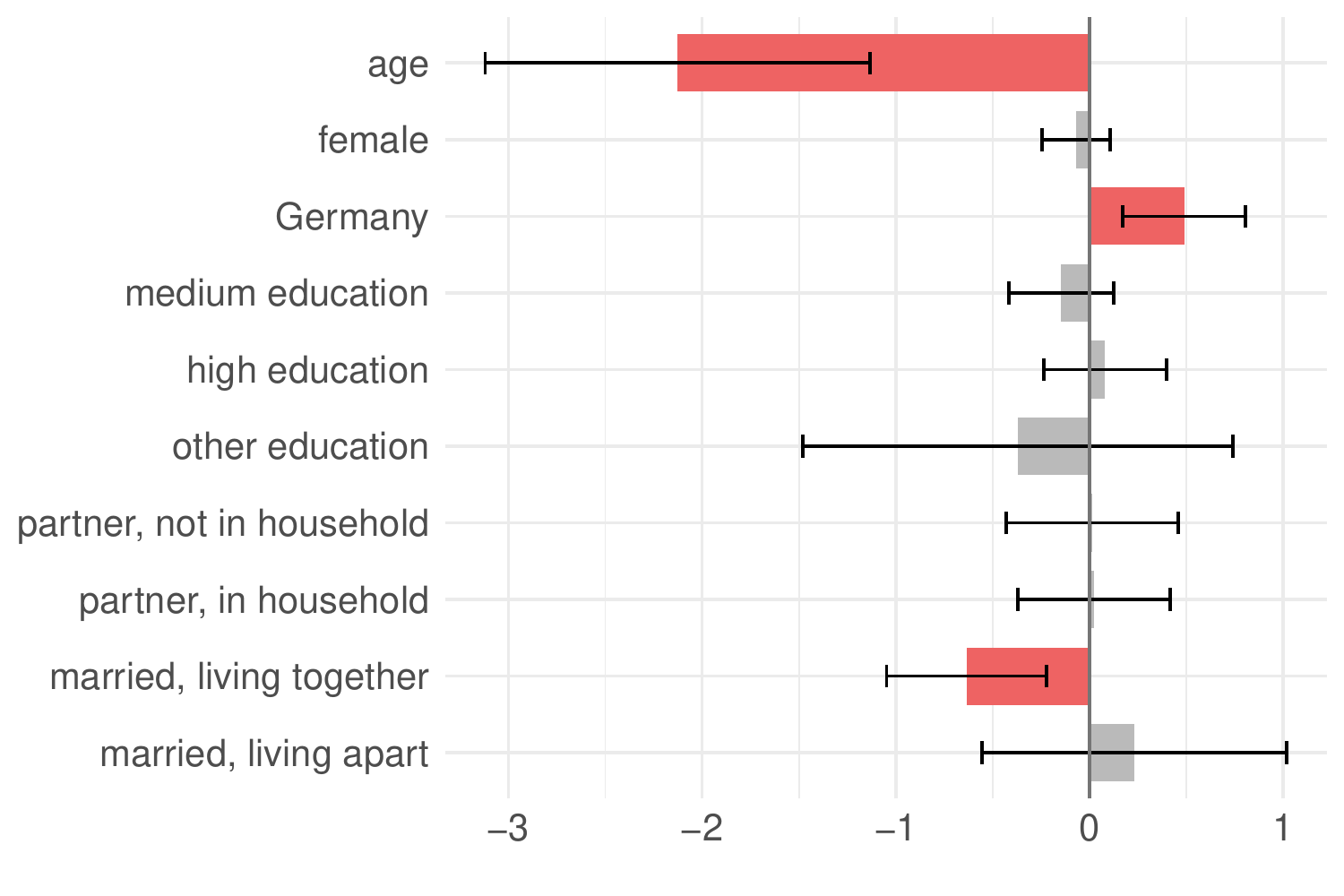}
\caption{Regression coefficients of the socio-demographic characteristics in the logistic regression model.}
\label{fig_socio}
\end{figure}

\begin{figure}[h]
\centering
\includegraphics[scale=1]{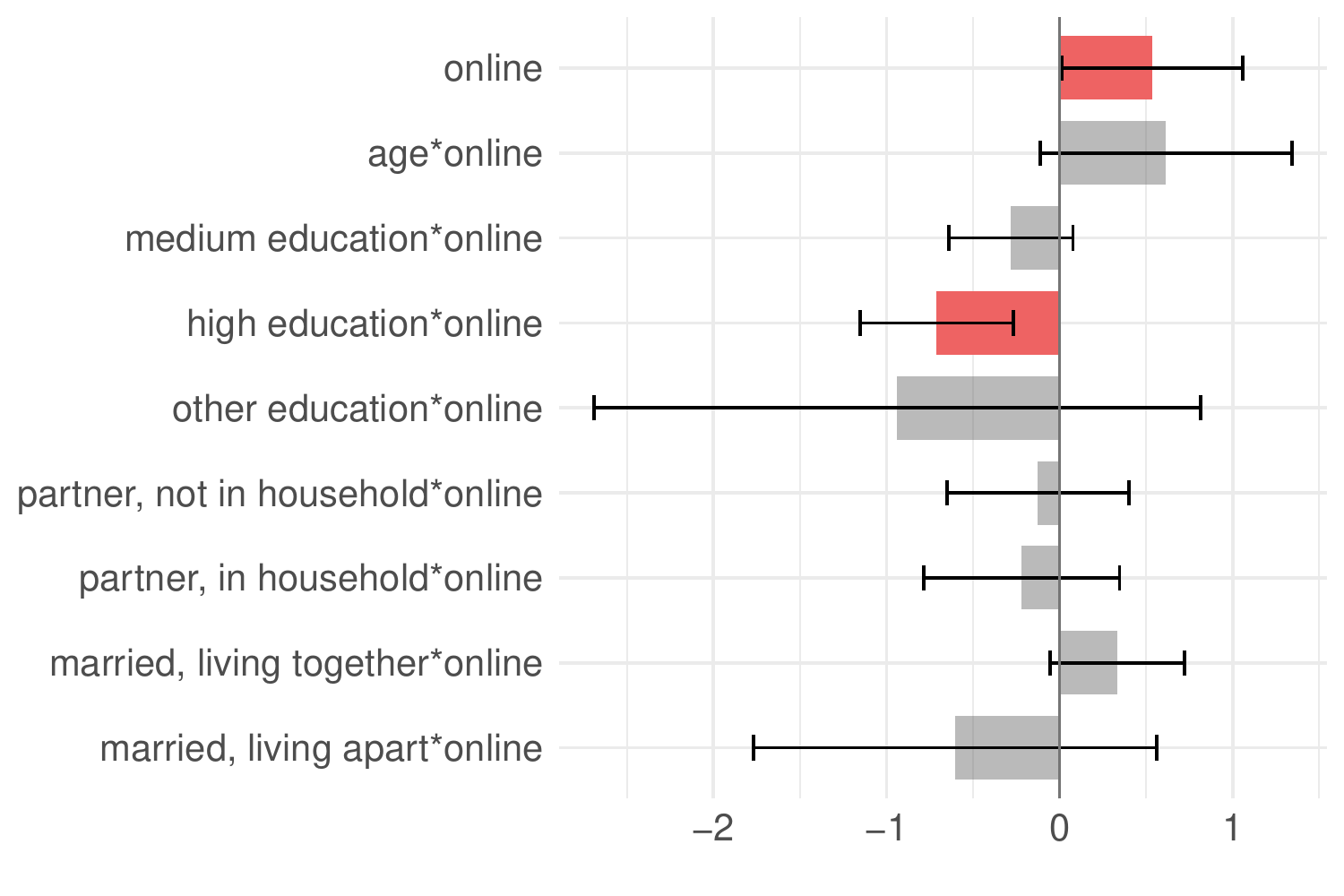}
\caption{Regression coefficients of the chosen survey mode in the logistic regression model.}
\label{fig_online}
\end{figure}

We do not find a significant effect for respondents' gender but do find a positive effect for having German citizenship: respondents with German citizenship dropped out after the recruitment survey at a higher rate than respondents without German citizenship. 

We find that survey mode interacts with age, education (though only significantly with high education) and living situation (only being significant at the $10\%$ level for ``married, living together''). We interpret the effects of all variables that show a significant interaction with chosen survey mode. The online mode has a positive effect and is positively interacted with age, which itself has a negative effect: the older the respondents, the lower their likelihood to drop out after the recruitment survey. The effect is much stronger for respondents who chose the paper-and-pencil mode $(-2.126)$ than those who chose the online mode $(-1.511)$. 
Having medium, high or other education is negatively interacted with online survey mode. Medium and other education both have negative main effects and negative (though not significant) interactions with online mode. Drop-out was lower for these two groups than for respondents with low education and the decreasing effect is less pronounced for respondents who chose the paper-and-pencil mode than it is for those who chose the online mode.
For high education, we find a positive effect on drop-out for respondents who chose the paper-and-pencil questionnaire $(0.081)$; this turns into a negative effect for highly educated respondents who chose the online mode $(-0.630)$. 
We find positive but not significant effects for the living situations ``not married with partner, separate households'', ``not married with partner, joint household'' and ``married, living apart'' and negative interactions with online mode for these categories. This means that, compared to respondents who were not married and did not have a partner, the risk of drop-out was higher for respondents who chose the paper-and-pencil mode but lower for those who chose the online mode. Compared to respondents who were not married and did not have a partner, respondents who were married and lived together with their spouse showed a significant reduction in drop-out after the recruitment interview that was stronger if they chose the paper-and-pencil questionnaire $(-0.634)$ than if they chose the online mode $(-0.300)$.


\subsection{Discussion of results}

We find that socio-demographic characteristics, survey mode, one measure of cooperativeness, and a rating of willingness to participate in the welcome survey explain survey drop-out after the recruitment survey. Losing respondents at this stage is not only very costly but can, through selective nonresponse, put the validity of panel inference at risk. Thus, the goal of panel recruitment should be to prevent panel drop-off  among population groups that are found to be least likely to become panel members. Knowing which population groups are likely to drop-out can  help in the identification and development of targeted strategies for these groups \citep{lynn2017standardised}. Especially interesting in this context is the moderating effect of mode choice. Knowing this, it might be worthwhile to develop targeted interventions that increase response depending on the mode the respondents choose. Further research is needed to determine which interventions are the most successful for different population groups.

\section{Conclusion}\label{sec4}
In this paper, we introduce double machine learning methods to survey statistics, enabling researchers to study causal relationships between treatment variables and dependent variables in high-dimensional data sets while controlling for large numbers of variables. In an application, we analyze drop-outs in the recruitment to the GESIS panel using double machine learning for logit regression. Classical machine learning is well suited to  \emph{predict} who will not respond to a survey but leads to biased estimates of causal relationships and invalid inference results and should therefore not be applied in studies aiming to \emph{explain} the effects of treatment variables in a high-dimensional setting. Performing valid post-selection inference, de-biased/double machine learning allows  the significant variables influencing the dependent variable to be identified. Unbiased estimation is crucial for learning (a) which treatments affect which dependent variables in which ways and (b) which factors to manipulate to achieve better outcomes. Given that survey scientists are confronted with many types of big data, such as  paradata from the web, or data from sensor tracking or mobile apps, the applications for which survey scientists might benefit from the double machine learning technique are numerous. For future research, we intend to analyze how double machine learning might be used to select and include high numbers of control variables in imputation or weighting models.


\pagebreak

\appendix

\section{Data and Empirical Results}
\begin{table}[ht]
\centering
\tiny
\begin{tabular}{rrrrrr}
  \hline
 & Coefficient & p-value & 2.5\% & 97.5\% \\ 
  \hline
Age & -2.126 & 0.000 & -3.120 & -1.133  \\ 
 Female & -0.069 & 0.440 & -0.246 & 0.107 \\ 
  Germany & 0.489 & 0.003 & 0.171 & 0.807  \\ 
  Good willingness to answer questions & -0.354 & 0.013 & -0.632 & -0.076  \\ 
  Rather easy to persuade respondent (interview) & -0.147 & 0.178 & -0.360 & 0.067  \\ 
  Very easy to persuade respondent (interview) & -0.209 & 0.111 & -0.465 & 0.048  \\ 
  Rather easy to persuade respondent (follow-up interview) & -0.114 & 0.353 & -0.354 & 0.127  \\ 
  Very easy to persuade respondent (follow-up interview) & -0.158 & 0.295 & -0.454 & 0.138 \\ 
  Rather likely to participate & -0.039 & 0.821 & -0.373 & 0.296  \\ 
  Very likely to participate & -0.436 & 0.012 & -0.775 & -0.098  \\ 
	Medium education & -0.146 & 0.289 & -0.417 & 0.124  \\ 
  High education & 0.081 & 0.616 & -0.235 & 0.397  \\ 
  Other education & -0.370 & 0.514 & -1.481 & 0.741  \\ 
	Not married with partner, separate households & 0.014 & 0.951 & -0.430 & 0.458  \\ 
  Not married with partner, joint household & 0.023 & 0.910 & -0.371 & 0.416  \\ 
  Married living together & -0.634 & 0.003 & -1.047 & -0.221  \\ 
  Married living apart & 0.233 & 0.562 & -0.554 & 1.019  \\ 
  Online & 0.537 & 0.044 & 0.015 & 1.058  \\ 
	Age*online & 0.615 & 0.098 & -0.114 & 1.343  \\ 
  Medium education*online & -0.282 & 0.123 & -0.640 & 0.077  \\ 
 High education*online & -0.711 & 0.002 & -1.153 & -0.269  \\ 
  Other education*online & -0.938 & 0.294 & -2.691 & 0.814  \\ 
  Not married with partner, separate households*online & -0.126 & 0.637 & -0.652 & 0.399  \\ 
  Not married with partner, joint household*online & -0.218 & 0.449 & -0.783 & 0.347  \\ 
  Married living together*online & 0.334 & 0.092 & -0.054 & 0.722  \\ 
  Married living apart*online & -0.605 & 0.309 & -1.770 & 0.560  \\ 
   \hline
\end{tabular}
\caption{Estimated treatment effects of the double lasso for logistic regression including p-values and confidence intervals.} 
\label{treatment_effects}
\end{table}

\begin{table}[!ht]
\centering
\tiny
\begin{tabular}{lll}
  \hline \hline 	  \\[0.8ex]
  Variable & Answer Categories & Code ($0$ is baseline) \\ [0.8ex]
  \hline \hline  \\ [-0.8ex]	
 Participation mode & Offline & 0: Offline \\
         & Online & 1: Online \\	
	 \hline	
Willingness of the respondent                       &  Good & 1: Good\\
 to answer the question            &  Medium & 0: Bad\\
                                   &  Bad & 0: Bad\\
																	 &  Good in the beginning but got worse & 0: Bad\\
																	 &  Bad in the beginning but got better & 0: Bad \\
 \hline																
Difficulty to persuade respondent & Very difficult & 0: Difficult\\
to take part in interview & Rather difficult& 0: Difficult\\
& Rather easy  &3: Rather easy\\ 
& Very easy & 4: Very easy\\
 \hline
Difficulty to persuade respondent   & Very difficult & 0: Difficult\\
to take part in follow-up interview & Rather difficult& 0: Difficult\\
& Rather easy & 3: Rather easy\\
& Very easy & 4: Very easy\\
 \hline		
Likelihood of participation in first & Very likely &  5: Very likely\\
 online- or paper questionnaire &  Rather likely& 2: Rather likely\\
 &Rather unlikely& 0: Unlikely\\
  &Very unlikely& 0: Unlikely\\
	 \hline		
Highest education & Still in school                                         & 8: High\\
                  & Left school without degree                              & 0: Low\\
									& Lower secondary degree                                  & 0: Low\\
									& Secondary degree                                        & 4: Medium\\
									& Polytechnical secondary degree (GDR) 8th or 9th grade   & 0: Low\\
									& Polytechnical secondary degree (GDR) 10th grade         & 4: Medium  \\
									& Advanced technical college certificate                  & 8: High\\
									& General qualification for university entrance           & 8: High\\
									& Other degree                                            & 9: Other\\
 \hline		
 Gender & Male & 0: Male \\
         & Female & 2: Female \\
				 \hline		
Citizenship & Germany & 0: Germany \\
            & EU28 & 4: Other \\
						& Rest of Europe & 4: Other \\
						& Other & 4: Other \\
						 \hline		
 Age            & Year of birth &  2013 - year of birth\\
\hline		
Living situation & Not married, no partner & 0: No partner \\
& Not married with partner, separate households & 1: Partner not in household\\
&  Not married with partner, joint household & 2: Partner in household\\
& Married living together & 3: Married living together \\
& Married living apart & 4: Married living apart\\
\hline  \hline	 \\[0.8ex]
\end{tabular} 
\caption{Answer categories and coding of the treatment variables} 
\label{treatment_coding}
\end{table}\ \\ 

\pagebreak

\section{Double Machine Learning for Logistic Regression}\label{sec_dml}

The double machine learning approach for logistic regression includes three main steps:
\begin{itemize} 
\item[(1)] initial estimation of the  regression function via post-lasso logistic regression, 
\item[(2)] estimation of instruments that are orthogonal to the weighted controls via weighted post-selection least squares, and 
\item[(3)] estimation of $\alpha_0$ based on the nuisance estimates obtained in (1) and (2). 
\end{itemize}

\noindent
The estimation procedure for $\alpha_0$ is summarized in more detail in the following algorithm:

\begin{algorithm}[H]
	\caption{DML for Logistic Regression} 
	\label{dmlalg}
	\begin{algorithmic}[1]
	
	\State Run a post-lasso-logistic regression of $Y_i$ on $D_i$ and $X_i$:
	\begin{align*}
	(\hat{\alpha}, \hat{\beta}) &\in \arg \min_{\alpha, \beta} \mathbb{E}_n[\Lambda_i(\alpha,\beta)] +\lambda_1/n ||(\alpha, \beta)||_1,\\
	(\tilde{\alpha}, \tilde{\beta}) &\in \arg \min_{\alpha, \beta} \mathbb{E}_n[\Lambda_i(\alpha,\beta)]: support(\beta) \subset support(\tilde{\beta}).
	\end{align*}
	For $i= 1,\ldots,n$, keep the value $X_i' \tilde{\beta}$ and weight $\hat{f}_i:=\hat{w_i}/\hat{\sigma}_i$, where 
	$$\hat{w}_i=G'(D_i \tilde{\alpha})+ X_i'\tilde{\beta}$$
	and
	$$\hat{\sigma}^2_i=Var(Y_i|D_i,X_i)=G(D_i \tilde{\alpha} + X_i' \tilde{\beta})\{1-G(D_i \tilde{\alpha} + X_i' \tilde{\beta})\}.$$
	
	\State Run a post-lasso OLS regression of $\hat{f}_i D_i$ on $\hat{f}_i X_i:$
	\begin{align*}
	\hat{\theta} &\in \arg \min_{\theta} \mathbb{E}_n[\hat{f}^2_i(D_i - X'_i \theta)^2]  + \lambda_2/ n ||\Gamma \theta||_1,\\
	\tilde{\theta} &\in \arg \min_{\theta} \mathbb{E}_n[\hat{f}^2_i(D_i  - X'_i \theta)^2]: support(\theta) \subset support(\hat{\theta}).
	\end{align*}
  Keep the residual $\hat{v}_i:=\hat{f}_i(D_i - X'_i \tilde{\theta}$ and instrument $\hat{z}_i:=\hat{v}_i/\sqrt{\hat{\sigma}_i}$, $i= 1,...,n$.
	
	\State Run an instrumental Logistic regression of $Y_i - X'_i \tilde{\beta}$ on $D_i$ using $\hat{z}_i$ as the instrument for $D_i$
	\begin{align*}
	\check{\alpha} \in \arg \inf_{\alpha\in\mathcal{A}} L_n(\alpha),
	\end{align*}
	where 
	$$L_n(\alpha) = \frac{|\mathbb{E}_n[\{Y_i - G(D_i \alpha+ X'_i \tilde{\beta})\}z_i]|^2}{\mathbb{E}_n[\{Y_i-G(D_i \alpha +X'_i \tilde{\beta})\}^2 \hat{z}^2_i]}$$
	and $\mathcal{A}=\{\alpha \in \mathbb{R}:|\alpha - \tilde{\alpha}| \leq C/ \log n \}.$ Compute the confidence region with asymptotic coverage $1-\xi$:
	\begin{align*}
	\{\alpha \in \mathbb{R}:|\alpha - \tilde{\alpha}| \leq \hat{\Sigma}_n \Phi^{-1}(1- \xi/2)/\sqrt{n}\}.
	\end{align*}
\end{algorithmic} 
\end{algorithm}
\noindent
For the estimator of the variance and details about the penalty levels, we refer to \cite{BCW2016}.

\newpage
\bibliographystyle{chicago}
\bibliography{Literatur_NR, mybibAR}

\begin{thebibliography}{}

\bibitem[\protect\citeauthoryear{Bach, Chernozhukov, and Spindler}{Bach
  et~al.}{2018}]{Bach2018}
Bach, P., V.~Chernozhukov, and M.~Spindler (2018, September).
\newblock {Valid Simultaneous Inference in High-Dimensional Settings (with the
  hdm package for R)}.
\newblock Papers 1809.04951, arXiv.org.

\bibitem[\protect\citeauthoryear{Belloni, Chernozhukov, Fern{\'a}ndez-Val, and
  Hansen}{Belloni et~al.}{2017}]{belloni2017program}
Belloni, A., V.~Chernozhukov, I.~Fern{\'a}ndez-Val, and C.~Hansen (2017).
\newblock Program evaluation and causal inference with high-dimensional data.
\newblock {\em Econometrica\/}~{\em 85\/}(1), 233--298.

\bibitem[\protect\citeauthoryear{Belloni, Chernozhukov, and Hansen}{Belloni
  et~al.}{2014a}]{BCH2014}
Belloni, A., V.~Chernozhukov, and C.~Hansen (2014a).
\newblock High-dimensional methods and inference on structural and treatment
  effects.
\newblock {\em Journal of Economic Perspectives\/}~{\em 28\/}(2), 29--50.

\bibitem[\protect\citeauthoryear{Belloni, Chernozhukov, and Hansen}{Belloni
  et~al.}{2014b}]{BelloniChernozhukovHansen2011}
Belloni, A., V.~Chernozhukov, and C.~Hansen (2014b).
\newblock Inference on treatment effects after selection amongst
  high-dimensional controls.
\newblock {\em Review of Economic Studies\/}~{\em 81}, 608--650.

\bibitem[\protect\citeauthoryear{Belloni, Chernozhukov, and Wei}{Belloni
  et~al.}{2013}]{belloni2013honest}
Belloni, A., V.~Chernozhukov, and Y.~Wei (2013).
\newblock Honest confidence regions for a regression parameter in logistic
  regression with a large number of controls.
\newblock cemmap working paper CWP67/13, London.

\bibitem[\protect\citeauthoryear{Belloni, Chernozhukov, and Wei}{Belloni
  et~al.}{2016}]{BCW2016}
Belloni, A., V.~Chernozhukov, and Y.~Wei (2016).
\newblock Post-selection inference for generalized linear models with many
  controls.
\newblock {\em Journal of Business \& Economic Statistics\/}~{\em 34\/}(4),
  606--619.

\bibitem[\protect\citeauthoryear{Blom, Bosnjak, Cornilleau, Cousteaux, Das,
  Douhou, and Krieger}{Blom et~al.}{2016}]{blom2016comparison}
Blom, A.~G., M.~Bosnjak, A.~Cornilleau, A.-S. Cousteaux, M.~Das, S.~Douhou, and
  U.~Krieger (2016).
\newblock A comparison of four probability-based online and mixed-mode panels
  in europe.
\newblock {\em Social Science Computer Review\/}~{\em 34\/}(1), 8--25.

\bibitem[\protect\citeauthoryear{Bosnjak, Dannwolf, Enderle, Schaurer,
  Struminskaya, Tanner, and Weyandt}{Bosnjak
  et~al.}{2018}]{bosnjak2018establishing}
Bosnjak, M., T.~Dannwolf, T.~Enderle, I.~Schaurer, B.~Struminskaya, A.~Tanner,
  and K.~W. Weyandt (2018).
\newblock Establishing an open probability-based mixed-mode panel of the
  general population in germany: The gesis panel.
\newblock {\em Social Science Computer Review\/}~{\em 36\/}(1), 103--115.

\bibitem[\protect\citeauthoryear{Breiman}{Breiman}{2001}]{breiman2001statistical}
Breiman, L. (2001).
\newblock Statistical modeling: The two cultures (with comments and a rejoinder
  by the author).
\newblock {\em Statistical science\/}~{\em 16\/}(3), 199--231.

\bibitem[\protect\citeauthoryear{Buskirk}{Buskirk}{2018}]{Buskirk2018}
Buskirk, T.~D. (2018).
\newblock Surveying the forests and sampling the trees: An overview of
  classification and regression trees and random forests with applications in
  survey research.
\newblock {\em Survey Practice\/}~{\em 11(1)}, 1--13.

\bibitem[\protect\citeauthoryear{Buskirk, Kirchner, Eck, and Signorino}{Buskirk
  et~al.}{2018}]{Buskirk2018_intro}
Buskirk, T.~D., A.~Kirchner, A.~Eck, and C.~S. Signorino (2018).
\newblock An introduction to machine learning methods for survey researchers.
\newblock {\em Survey Practice\/}~{\em 11(1)}, 1--10.

\bibitem[\protect\citeauthoryear{Buskirk and Kolenikov}{Buskirk and
  Kolenikov}{2015}]{buskirk2015finding}
Buskirk, T.~D. and S.~Kolenikov (2015).
\newblock Finding respondents in the forest: A comparison of logistic
  regression and random forest models for response propensity weighting and
  stratification.
\newblock {\em Survey Methods: Insights from the Field\/}, 1--17.

\bibitem[\protect\citeauthoryear{Callegaro}{Callegaro}{2013}]{callegaro2013paradata}
Callegaro, M. (2013).
\newblock Paradata in web surveys.
\newblock In F.~Kreuter (Ed.), {\em Improving surveys with paradata: Analytic
  uses of process information}, Chapter~11, pp.\  261--279. Wiley Online
  Library.

\bibitem[\protect\citeauthoryear{Chernozhukov, Chetverikov, Demirer, Duflo,
  Hansen, Newey, and Robins}{Chernozhukov
  et~al.}{2018}]{chernozhukov2018double}
Chernozhukov, V., D.~Chetverikov, M.~Demirer, E.~Duflo, C.~Hansen, W.~Newey,
  and J.~Robins (2018, 01).
\newblock {Double/debiased machine learning for treatment and structural
  parameters}.
\newblock {\em The Econometrics Journal\/}~{\em 21\/}(1), C1--C68.

\bibitem[\protect\citeauthoryear{Chernozhukov, Hansen, and
  Spindler}{Chernozhukov et~al.}{2015}]{CHS:2016}
Chernozhukov, V., C.~Hansen, and M.~Spindler (2015).
\newblock Valid post-selection and post-regularization inference: An
  elementary, general approach.
\newblock {\em Annual Review of Economics\/}~{\em 7\/}(1), 649--688.

\bibitem[\protect\citeauthoryear{Durrant and Steele}{Durrant and
  Steele}{2009}]{durrant2009multilevel}
Durrant, G.~B. and F.~Steele (2009).
\newblock Multilevel modelling of refusal and non-contact in household surveys:
  evidence from six uk government surveys.
\newblock {\em Journal of the Royal Statistical Society: Series A (Statistics
  in Society)\/}~{\em 172\/}(2), 361--381.

\bibitem[\protect\citeauthoryear{Earp, Mitchell, McCarthy, and Kreuter}{Earp
  et~al.}{2014}]{earp2014modeling}
Earp, M., M.~Mitchell, J.~McCarthy, and F.~Kreuter (2014).
\newblock Modeling nonresponse in establishment surveys: Using an ensemble tree
  model to create nonresponse propensity scores and detect potential bias in an
  agricultural survey.
\newblock {\em Journal of Official Statistics\/}~{\em 30\/}(4), 701--719.

\bibitem[\protect\citeauthoryear{Eck}{Eck}{2018}]{Adam2018}
Eck, A. (2018).
\newblock Neural networks for survey researchers.
\newblock {\em Survey Practice\/}~{\em 11(1)}, 1--10.

\bibitem[\protect\citeauthoryear{Feddersen, Metcalfe, and Wooden}{Feddersen
  et~al.}{2016}]{feddersen2016subjective}
Feddersen, J., R.~Metcalfe, and M.~Wooden (2016).
\newblock Subjective wellbeing: why weather matters.
\newblock {\em Journal of the Royal Statistical Society: Series A (Statistics
  in Society)\/}~{\em 179\/}(1), 203--228.

\bibitem[\protect\citeauthoryear{GESIS}{GESIS}{2020}]{ZA5665}
GESIS (2020).
\newblock Gesis panel - standard edition.
\newblock GESIS Datenarchiv, K{\"o}ln. ZA5665 Datenfile Version 37.0.0,
  https://doi.org/10.4232/1.13573.

\bibitem[\protect\citeauthoryear{Japec, Kreuter, Berg, Biemer, Decker, Lampe,
  Lane, O’Neil, and Usher}{Japec et~al.}{2015}]{10.1093/poq/nfv039}
Japec, L., F.~Kreuter, M.~Berg, P.~Biemer, P.~Decker, C.~Lampe, J.~Lane,
  C.~O’Neil, and A.~Usher (2015, 11).
\newblock {Big Data in Survey Research: AAPOR Task Force Report}.
\newblock {\em Public Opinion Quarterly\/}~{\em 79\/}(4), 839--880.

\bibitem[\protect\citeauthoryear{Kern, Klausch, and Kreuter}{Kern
  et~al.}{2019}]{kern2019tree}
Kern, C., T.~Klausch, and F.~Kreuter (2019).
\newblock Tree-based machine learning methods for survey research.
\newblock {\em Survey Research Methods\/}~{\em 13\/}(1), 73--93.

\bibitem[\protect\citeauthoryear{Kern, Weiss, and Kolb}{Kern
  et~al.}{2019}]{kern2019longitudinal}
Kern, C., B.~Weiss, and J.-P. Kolb (2019).
\newblock A longitudinal framework for predicting nonresponse in panel surveys.
\newblock Papers 1909.13361, arXiv.org.

\bibitem[\protect\citeauthoryear{Kirchner and Signorino}{Kirchner and
  Signorino}{2018}]{Kirchner2018}
Kirchner, A. and C.~S. Signorino (2018).
\newblock Using support vector machines for survey research.
\newblock {\em Survey Practice\/}~{\em 11(1)}, 1--11.

\bibitem[\protect\citeauthoryear{Kreuter}{Kreuter}{2013a}]{kreuter2013facing}
Kreuter, F. (2013a).
\newblock Facing the nonresponse challenge.
\newblock {\em The ANNALS of the American Academy of Political and Social
  Science\/}~{\em 645\/}(1), 23--35.

\bibitem[\protect\citeauthoryear{Kreuter}{Kreuter}{2013b}]{kreuter2013improving}
Kreuter, F. (2013b).
\newblock {\em Improving surveys with paradata: Analytic uses of process
  information}, Volume 581.
\newblock John Wiley \& Sons.

\bibitem[\protect\citeauthoryear{Liu}{Liu}{2020}]{liu2020using}
Liu, M. (2020).
\newblock Using machine learning models to predict attrition in a survey panel.
\newblock {\em Big Data Meets Survey Science: A Collection of Innovative
  Methods\/}, 415--433.

\bibitem[\protect\citeauthoryear{Lynn}{Lynn}{2017}]{lynn2017standardised}
Lynn, P. (2017).
\newblock From standardised to targeted survey procedures for tackling
  non-response and attrition.
\newblock {\em Survey Research Methods\/}~{\em 11\/}(1), 93--103.

\bibitem[\protect\citeauthoryear{Lynn}{Lynn}{2020}]{lynn2020}
Lynn, P. (2020).
\newblock Methods for recruitment and retention.
\newblock {\em Understanding Society Working Paper 2020-07\/}.

\bibitem[\protect\citeauthoryear{Murphy, Link, Childs, Tesfaye, Dean, Stern,
  Pasek, Cohen, Callegaro, and Harwood}{Murphy et~al.}{2014}]{murphy2014social}
Murphy, J., M.~W. Link, J.~H. Childs, C.~L. Tesfaye, E.~Dean, M.~Stern,
  J.~Pasek, J.~Cohen, M.~Callegaro, and P.~Harwood (2014, 11).
\newblock {Social Media in Public Opinion Research: Executive Summary of the
  Aapor Task Force on Emerging Technologies in Public Opinion Research}.
\newblock {\em Public Opinion Quarterly\/}~{\em 78\/}(4), 788--794.

\bibitem[\protect\citeauthoryear{Phipps, Toth, et~al.}{Phipps
  et~al.}{2012}]{phipps2012analyzing}
Phipps, P., D.~Toth, et~al. (2012).
\newblock Analyzing establishment nonresponse using an interpretable regression
  tree model with linked administrative data.
\newblock {\em The Annals of Applied Statistics\/}~{\em 6\/}(2), 772--794.

\bibitem[\protect\citeauthoryear{Plewis, Calderwood, and Mostafa}{Plewis
  et~al.}{2017}]{Plewis2017can}
Plewis, I., L.~Calderwood, and T.~Mostafa (2017).
\newblock Can interviewer observations of the interview predict future
  response?
\newblock {\em Methods, data, analyses\/}~{\em 11\/}(1).

\bibitem[\protect\citeauthoryear{Ro{\ss}mann and Gummer}{Ro{\ss}mann and
  Gummer}{2016}]{rossmann2016using}
Ro{\ss}mann, J. and T.~Gummer (2016).
\newblock Using paradata to predict and correct for panel attrition.
\newblock {\em Social Science Computer Review\/}~{\em 34\/}(3), 312--332.

\bibitem[\protect\citeauthoryear{Sakshaug and Huber}{Sakshaug and
  Huber}{2015}]{Sakshaug2020}
Sakshaug, J.~W. and M.~Huber (2015, 10).
\newblock An evaluation of panel nonresponse and linkage consent bias in a
  survey of employees in germany.
\newblock {\em Journal of Survey Statistics and Methodology\/}~{\em 4\/}(1),
  71--93.

\bibitem[\protect\citeauthoryear{Shmueli}{Shmueli}{2010}]{shmueli2010explain}
Shmueli, G. (2010).
\newblock To explain or to predict?
\newblock {\em Statistical science\/}~{\em 25\/}(3), 289--310.

\bibitem[\protect\citeauthoryear{Signorino and Kirchner}{Signorino and
  Kirchner}{2018}]{Signorino2018}
Signorino, C.~S. and A.~Kirchner (2018).
\newblock Using lasso to model interactions and nonlinearities in survey data.
\newblock {\em Survey Practice\/}~{\em 11(1)}, 1--10.

\bibitem[\protect\citeauthoryear{Sinibaldi and Eckman}{Sinibaldi and
  Eckman}{2015}]{sinibaldi2015using}
Sinibaldi, J. and S.~Eckman (2015).
\newblock Using call-level interviewer observations to improve response
  propensity models.
\newblock {\em Public Opinion Quarterly\/}~{\em 79\/}(4), 976--993.

\bibitem[\protect\citeauthoryear{Struminskaya, Lugtig, Keusch, and
  H{\"o}hne}{Struminskaya et~al.}{2020}]{struminskaya2020augmenting}
Struminskaya, B., P.~Lugtig, F.~Keusch, and J.~K. H{\"o}hne (2020).
\newblock Augmenting surveys with data from sensors and apps: Opportunities and
  challenges.
\newblock {\em Social Science Computer Review\/}, 1 -- 13.

\bibitem[\protect\citeauthoryear{Zinn and Gnambs}{Zinn and
  Gnambs}{2020}]{Zinn2020}
Zinn, S. and T.~Gnambs (2020).
\newblock Analyzing nonresponse in longitudinal surveys using bayesian additive
  regression trees: A nonparametric event history analysis.
\newblock {\em Social Science Computer Review\/}, 1 -- 22.

\end{thebibliography}

\end{document}